\documentclass[pre,preprint,showpacs]{revtex4-1}

\usepackage{graphicx}

\usepackage{amsmath,amssymb}
\usepackage{bm}
\usepackage{cancel}
\usepackage{subfigure} 
\usepackage{bbold}

\begin{document}
\title{Critical radius and temperature for buckling in graphene}
\author{ L.\ L.\ Bonilla$^1$, M.Ruiz-Garcia$^1$}
\affiliation {$^1$G. Mill\'an Institute, Fluid Dynamics, Nanoscience
and Industrial Mathematics, Universidad Carlos III de Madrid, 28911
Legan\'es, Spain}

\date{\today}
\begin{abstract} 
In this work, we find an analytical flat-membrane solution to the saddle point equations, derived by Guinea \emph{et al.} [Phys. Rev. B {\bf 89}, 125428 (2014)], for the case of a suspended graphene membrane of circular shape. We also find how different buckled membrane solutions bifurcate from the flat membrane at critical temperatures and membrane radii. The saddle point equations take into account electron-phonon coupling and this coupling provides a residual stress even for a flat graphene layer. Below a critical temperature (which is exceedingly high for an infinite layer) or above a critical size that depend on boundary conditions, different buckling modes that may be the germ of rippling appear. Our results provide the opportunity to develop new feasible experiments dealing with buckling in small suspended graphene membranes that could verify them. These experiments may also be used to fit the phonon-electron coupling constant or the bending energy.
\end{abstract}

\maketitle

\section{Introduction}

The discovery of graphene has spawned one of the most fertile scientific fields of today. Its unique electrical \cite{nov05} and mechanical properties \cite{lee08} promise to revolutionize current technology, making extremely important to characterize the properties of graphene in order to design and optimize new devices. Among mechanical properties, the presence of ripples in suspended graphene \cite{mey07} has aroused a great theoretical effort focused on their explanation from first principles \cite{fas07,abe07,kim08,gazit09,san-jose,gui14,gonzalez}, and by using simple statistical mechanics models \cite{jsm12bon,pre12bon,Ruiz-Garcia,prb12bon,schoelz15}. Thermal and external stresses have been used to create large ripples (wrinkles) in a layer of graphene which is suspended over trenches \cite{bao09}. These ripples have been explained using classical elasticity \cite{bao09} and seem to be qualitatively different from the nanometer sized ripples observed in \cite{mey07,kiri,locatelli}. The latter tend to be considered inherent to suspended graphene. Even when ripples in graphene are not fully understood, one of the most promising approaches to ripple formation is the quantum mechanical study of the membrane and the phonon-electron interaction as a mechanism for rippling \cite{gazit09,san-jose}. Since the graphene Debye temperature is $T_D\sim 1000$ K, the quantum treatment of graphene seems justified at room temperature. Perturbation treatments point to the vanishing of the renormalized bending energy of the membrane as a possible mechanism for ripple formation \cite{san-jose,gonzalez}. The destabilizing effects of quantum fluctuations on the bending rigidity of crystalline membranes (without electron-phonon interaction) have been considered in \cite{kats}. Within equilibrium theory and without phonon-electron interaction, the effect of (Matsubara) frequency-dependent renormalization of the anharmonic coupling after elimination of in-plane phonons has been considered in \cite{amo14}.

Recent experimental studies emphasize buckling effects in suspended graphene, as buckling is more easily characterized than ripple formation and can strongly affect the design of new graphene-based devices. Long-range buckling of the graphene sheet is pronounced near defects and dislocations \cite{SN88,CN93,leh13,kot14,rob14,BCGW15}. In \cite{lin12}, buckled bilayer and monolayer graphene are fully clamped to circular holes in an electrode. Buckling is induced by the fabrication process so that the sheets are buckled and strain-free at zero applied electric field. In the experiment, an external electric field pressures the membrane and induces a sudden change in the buckling direction (a ``snap-through'' effect). A relation between the critical pressure and the bending rigidity of bilayer graphene provides a measure of the latter. Even when the graphene layer is not buckled initially, a strong enough electrostatic force can produce irreversible buckling \cite{svensson11}. A spontaneous buckling effect (``mirror buckling'') is also found by Xu \emph{et al.} in \cite{xu14}. This effect is systematically studied in \cite{schoelz15}, in a complementary approach to that in \cite{lin12}. In \cite{schoelz15}, the authors study a suspended graphene monolayer using scanning tunnelling microscopy (STM). The microscope tip is located on the centre of an initially non-buckled layer. The STM keeps a constant current and a variable potential between the tip and the sample. Once the current is fixed at a sufficiently high value, the sample buckles when the potential increases, similarly to the experiments in \cite{lin12,svensson11}. 

These experimental efforts have their theoretical counterpart through different approaches to the buckling of two dimensional (2D) layers such as graphene. A broad overview of the effects of strain in graphene and its relation with its electrical properties can be found in \cite{amorim15}. The study of spontaneous buckling in graphene due to doping is carried out in \cite{gazit09}. In polymerized membranes, buckling occurs below a critical temperature \cite{Guitter88}. Phenomenological models of crystalline membranes coupled to spins undergoing Glauber dynamics also indicate that there are critical temperatures below which the membranes buckle \cite{jsm12bon, Ruiz-Garcia}.

In \cite{gui14}, Guinea \emph{et al.} have developed a model for suspended graphene based in the coupling between flexural phonons and electrons. They consider a 2D membrane in equilibrium that is embedded in a $2+d$ dimension space, where there are $d$ dimensional \emph{out-of-membrane} displacements ($d=1$ in the physical situation). Then they eliminate the in-plane phonons, ignore the frequency dependence of the resulting couplings, use the self-consistent screening approximation and perform a saddle-point analysis of the free energy in the limit as $d\to\infty$. The resulting saddle-point equations (SPEs) are time-independent and will be analyzed in the present paper. They consist of one von K\'arm\'an type plate equation for the physical out-of-plane displacement coupled to two equations for two auxiliary fields: one ``scalar'' stress associated to membrane curvature and a field associated to charge fluctuations. We expect these equations to describe buckling and ripples of the graphene membrane in a stable stationary (equilibrium) configuration after all possible transients have decayed. We find {\em flat membrane} solutions with constant non-zero auxiliary fields. The linearization of the {\em stationary} SPEs about these solutions accompanied by appropriate boundary conditions provide an eigenvalue problem that yields critical values of temperature and membrane size corresponding to the bifurcation of buckling states from the flat membrane. Why? 

The stationary SPEs provide stationary solutions to {\em not yet derived} dynamic SPEs that should describe the graphene membrane out of equilibrium. Among them, we have found the stationary flat membrane solution. Other solutions may bifurcate from it at appropriate parameter values. To find them, we have to solve the eigenvalue problem that governs the linear stability of the stationary flat membrane solution to dynamic SPEs. The corresponding eigenvalues give the growth of disturbances about the stationary solution. We do not know the precise shape of the dynamic SPEs but we may surmise that stationary solutions bifurcating from the flat membrane appear when these eigenvalues are zero. But linearized dynamic SPEs with zero eigenvalues are the same as linearized stationary SPEs, which we know from \cite{gui14}. These linearized stationary SPEs have nonzero solutions only for particular values of the constant auxiliary fields that correspond to critical temperature and membrane sizes. The eigenmodes solving these equations give the shape of the buckling states near bifurcation points. A combination of eigenmodes may characterize ripples in graphene.

The rest of the paper is as follows. In section \ref{SPE} we briefly revise the derivation of the stationary saddle-point equations \cite{gui14} and write them in real (not Fourier) space. In section \ref{constant_s_and_a}, we find the flat membrane solution of the SPEs for constant auxiliary fields. In Section \ref{eigenvalue}, we linearize the plate equation that forms part of the stationary SPE about the flat solution and add appropriate boundary conditions for a finite circular graphene sheet. Thus we obtain an eigenvalue problem for critical values of the constant auxiliary fields. From the critical auxiliary fields, we get critical values of temperature and membrane size at which non-flat buckled solutions may bifurcate from flat ones. The linearized plate equations are solved for a circular monolayer of graphene with two different boundary conditions corresponding to a free graphene layer and to a clamped sample. We obtain different eigenmodes (corresponding to vertical deformations of the layer) together with their respective critical temperatures and radii. The critical radii allow us to predict the minimal size that allows a graphene layer to buckle. Moreover the combination of the different bifurcating modes could be used to characterize ripples in graphene. The last section contains our conclusions.

\section{Model and saddle-point equations}
\label{SPE}
In \cite{gui14}, Guinea, Le Doussal and Wiese have considered the graphene sheet to be a 2D membrane embedded in a larger space of dimension $d+2$ and interacting with $N_fd$ copies of a free Dirac fermion. The real physical system has $N_f=4$ (four flavors, two valleys and two spins) and $d=1$ but Guinea {\em et al} have derived saddle-point equations involving the out-of-plane displacement in the limit as $d\to\infty$. Here we analyze buckling of the graphene membrane using the saddle-point equations. Let us briefly recall the model, the saddle-point equations and their meaning. The model Hamiltonian consists of:  
\begin{itemize}
\item The deformation energy of the graphene sheet is the sum of kinetic energy, $H_{kin}$, and of curvature and elastic energy, $H_{elas}$, with
\begin{equation}
H_{elas}=\frac{1}{2}\int d^2x[\kappa(\nabla^2h_a)^2+\lambda u_{ii}^2+2\mu u_{ij}^2].
\end{equation}
Here $u_i$, $i=1,2$, are the in-plane phonon displacements, $h_a$, $a=1,...,d$ are the out-of-plane flexural phonon modes, $u_{ij}:=\frac{1}{2}(\partial_i u_j +\partial_j u_i+\partial_ih_a\partial_jh_a)$, $\lambda$ and $\mu$ are the Lam\'e constants, and $\kappa$ is the bending energy.
\item The energy of the free Dirac fermions (in units such that $\hbar=1$),
\begin{equation}
H_e=\int d^2x \sum_{\gamma=1}^{N_f d}\bar{\Psi}_\gamma [-v_F\bm{\sigma}\cdot(-i\nabla)]\Psi_\gamma.
\end{equation}
Here $\bm{\sigma}=(\sigma_x,\sigma_y)$ are the Pauli matrices, $v_F$ is the Fermi velocity of Dirac electrons in graphene such that, in these units, $\frac{v_F}{a}=5$ eV ($a=1.4$ \AA \ is the side of a graphene hexagon).
\item The electron-phonon coupling
\begin{eqnarray}
H_{e-ph}=-g_0\int d^2x \delta\rho(x) u_{ii}(x),\\
\delta \rho(x)=\rho(x)-\rho_0=\frac{1}{d}\sum_{\gamma=1}^{N_f d}\bar{\Psi}_\gamma \mathbb{1}\Psi_\gamma -\rho_0,\label{deltarho}
\end{eqnarray}
where $\delta\rho(x)$ is the the charge distribution on the graphene layer, $\rho_0$ is the equilibrium carrier density, and $g_0$ is in the range between 4 and 50 eV. 
\item Finally the Coulomb interaction has the form
\begin{equation}
H_{ee}=\frac{1}{2} \int \frac{d^2q}{(2\pi)^2}V_0(q)|\rho(q)|^2,
\end{equation}
where $V_0(q)=\frac{2\pi e^2}{\epsilon_0 q}$ is the Fourier transform of the electrostatic potential, $V_0(r)=e^2/(\epsilon_0 r)$, $\epsilon_0$ is the dielectric constant of the environment and $-e<0$ is the charge of the electron.
\end{itemize}
 
Once all interactions are defined, the resultant Hamiltonian is integrated over the in-plane phonons, leading to a coupled theory of flexural phonon modes and electrons. The frequency dependence of the resulting couplings is ignored, as it is important only for temperatures below 90 K \cite{amo14}, that we do not consider here. In this process, the Coulomb interaction becomes
\begin{equation}
\hat{V}(q)=\frac{2\pi e^2}{\epsilon_0 q}-\frac{g_0^2}{\lambda+2\mu}.\label{eq7}
\end{equation}
The effective Hamiltonian, depending only on the electrons and flexural phonon modes, is used to build a Matsubara equilibrium partition function. For later convenience, this partition function is transformed using two fluctuating auxiliary fields, $\sigma$ and $\alpha$, defined through the relation,
\begin{eqnarray}
\left(\begin{array}{c}
\sigma(x)\\
\alpha(x)\\
\end{array}\right)=\int_{x'} \left(\begin{array}{cc}
K_0& -g\\
-g & \hat{V}\\
\end{array}\right)_{xx'}
\left(\begin{array}{c}
\Phi(x')\\
\delta\rho(x')\\
\end{array}\right)=
\left(\begin{array}{c}
K_0 \Phi(x) -g \delta\rho(x)\\
-g \Phi(x)+ \int_{x'} \hat{V}(x-x') \delta\rho(x')\\
\end{array}\right)\!.
\label{eqn1}
\end{eqnarray}
Here $g=\frac{2\mu}{2\mu +\lambda}g_0\sim g_0$, $K_0=4\mu(\mu+\lambda)/(2\mu+\lambda)d$, $(K_0)_{xx'}=K_0\delta(x-x')$, $(g)_{xx'}=g\delta(x-x')$ and table \ref{table3} gives the dimensions of parameters and variables. In \eqref{eqn1}, $\Phi(x)$ is  
\begin{eqnarray}
\Phi(x)=\frac{1}{d}\sum\limits_{a=1}^{d} \frac{1}{2} P^T_{ij}(\partial)\partial_i h_a \partial_j h_a,\label{phi}
\end{eqnarray}
where $P^T_{ij}=\delta_{ij}-\frac{\partial_i\partial_j}{\nabla^2}$ is the transversal projector. With this definition, $\Phi(x)$ is related to the Gaussian curvature ($\mathcal{K}$) through the relation $\Phi(q)=\mathcal{K}(q)/q^2$. In (\ref{eqn1}), $\sigma(x)$ is a local linear combination of  $\Phi(x)$ and the charge disturbance $\delta\rho(x)$, whereas $\alpha(x)$ is a nonlocal linear combination, as it involves the potential energy of one electron in $x$ with respect to the charge disturbance distribution in $x'$, $\int_{x'}\hat{V}(x-x')\delta\rho(x')$. The action corresponding to the Matsubara partition function is 
\begin{align}
S=&\int d^2x \int\limits_{0}^{\beta} d\tau \sum\limits_{a=1}^{d} \left[\frac{\rho}{2}(\partial_\tau h_a)^2+\frac{\kappa}{2}(\nabla^2h_a)^2\right]+\frac{1}{\beta}\sum\limits_{\omega'_n}\int_{q} \Bigg\{\sum\limits_{\gamma=1}^{N_fd} \bar{\Psi}_\gamma(-q,-\omega'_n) \nonumber\\
 &[-v_F \mathbf{\sigma}\cdot (-i\nabla) -(i\omega'_n+\mu)\mathbb{1}]\Psi_\gamma(q,\omega'_n)\Bigg\} +\int_{x\tau} \Bigg\{\sigma(x) \Big[\frac{1}{2} P^T_{ij}  (\partial)\sum\limits_{a=1}^{d} \partial_i h_a \partial_j h_a\Big] \nonumber\\
 &+\alpha(x)   \sum\limits_{\gamma=1}^{N_fd} \bar{\Psi}_\gamma \mathbb{1}\Psi_\gamma \Bigg\}-\frac{d}{2}\int\limits_{xx'\tau} 
 \left(\begin{array}{cc}
 \sigma&
 \alpha
 \end{array}\right)_{x\tau}   
 \left(\begin{array}{cc}
 K_0& -g\\
 -g & \hat{V}\\
 \end{array}\right)_{xx'}^{-1}
 \left(\begin{array}{c}
 \sigma\\
 \alpha\\
 \end{array}\right)_{x'\tau},
 \label{S}
\end{align}
where $\mu$ is the chemical potential. We now decompose the vertical displacements into an average and a fluctuating part, $h_a=\langle h_a \rangle + \delta h_a$, and assume symmetry breaking in the direction $a=1$, $\langle h_a \rangle=\delta_{a1}h_1 \neq 0$. Integrating over the fermions and the fluctuating part of the vertical displacements, the action is \cite{gui14}
\begin{align}
\frac{S'}{d}&=\frac{1}{2} \text{tr}\ln(-\rho \partial_\tau^2 +\kappa\nabla^4- \left[  P^T_{ij}(\partial)  \sigma(x,\tau)  \right] \partial_i\partial_j)-\frac{N_f}{2} \text{tr}\ln(-v_F \left[ \mathbf{\sigma} \cdot (-i\nabla)\right] +\left[ \alpha(x,\tau) -\mu -\partial_\tau\right] \mathbb{1})   \nonumber\\
&-\frac{1}{2}
\int\limits_{xx'\tau} 
 \left(\begin{array}{cc}
 \sigma
 &\alpha
 \end{array}\right)_{x\tau}   
 \left(\begin{array}{cc}
 K_0& -g\\
 -g & \hat{V}\\
 \end{array}\right)_{xx'}^{-1}
 \left(\begin{array}{c}
 \sigma\\
 \alpha\\
 \end{array}\right)_{x'\tau}
 +\frac{1}{d}\int\limits_{x,\tau} \left[ \frac{\kappa}{2} (\nabla^2h_1)^2+ \frac{\rho}{2}(\partial_\tau h_1)^2+\frac{\sigma}{2}P^T_{ij}(\partial)\partial_i h_1 \partial_j h_1  \right].
\label{S2}
\end{align}
\begin{table}[]
\centering
\begin{tabular}{|cccccccccc|}
  \hline
  \multicolumn{10}{|c|}{Dimensions of variables} \\
  \hline
 $\sigma(x)$&$\alpha(x)$&$K_0$&$g$&$V(x)$&$\hat{V}(q)$&$\Phi(x)$&$\delta\rho(x)$&$v_F$&$\kappa$  \\
 $\frac{E}{L^2}$&$E$&$\frac{E}{L^2}$&$E$&$E$&$EL^2$&$1$&$L^{-2}$&$EL$&$E$\\
 \hline
\end{tabular}
\caption{Dimensions of the parameters and variables appearing in equations (\ref{eqn1}), (\ref{eq1}) and (\protect \ref{eq2}). $E$ for energy and $L$ for length dimension.}
\label{table3}
\end{table}
To obtain the saddle-point equations, we vary (\ref{S2}) with respect to $\sigma$, $\alpha$ and $h_1$, thereby obtaining 
\begin{eqnarray}
&&\int_{x'} \left(\begin{array}{cc}
K_0& -g\\
-g & \hat{V}\\
\end{array}\right)^{-1}_{xx'} \left(\begin{array}{c}
\sigma_0(x')\\
\alpha_0(x')\\
\end{array}\right)=\nonumber\\
&& \left(\begin{array}{c}
P^T_{ij}\partial_{x_i}h(x)\partial_{x_j}h(x)-\frac{1}{2\beta}\sum_{\omega_n}P^T_{ij}\partial_{x_i}\partial_{x_j} [\rho\omega^2_n+\kappa\Delta_y^2-P^T_{lm}\sigma_0(y)\partial_{y_l}\partial_{y_m}]^{-1}_{xx} \\
\frac{4}{\beta}\sum_{\omega'_n}(i\omega'_n)[(i\omega'_n-\alpha_0(y)+\mu)^2+v_F^2\nabla_y^2]^{-1}_{xx} \\
\end{array}\right)\!. \label{eq1}
\end{eqnarray}
\begin{equation}
-\kappa\Delta^2h+\partial_{x_i}(\sigma_{ij}\partial_{x_j}h)=0,\label{eq2}
\end{equation}
\begin{equation}
\partial_{x_i}\sigma_{ij}=0,  \label{eq2b}
\end{equation}
\begin{equation}
\sigma_{ij}=P^T_{ij}\sigma_0=(\delta_{ij}-\partial_{x_i}\partial_{x_j}\Delta^{-1})\sigma_0, \label{eq2c}
\end{equation}
where we have replaced $h_1=h$ and not yet made $\mu=0$ as in \cite{gui14}. Eq. (\ref{eq1})  provides the average fields $\sigma_0=\langle\sigma(x)\rangle$ and $\alpha_0=\langle\alpha\rangle$  in terms of $h$. The average auxiliary fields are related to the averages of the charge distribution (\ref{deltarho}) and of the field (\ref{phi}), 
\begin{eqnarray}
\Phi_0(x)=\frac{1}{d}\left\langle\sum\limits_{a=1}^{d} \frac{1}{2} P^T_{ij}(\partial)\partial_i h_a \partial_j h_a\right\rangle\!,\quad \delta\rho_0(x)=\langle\delta\rho(x)\rangle,
\end{eqnarray}
by the linear equation (\ref{eqn1}), which also holds for the averages of the corresponding quantities. (\ref{eq2}), \eqref{eq2b} and \eqref{eq2c} are von Karman plate equations (they appear in Fourier transform form in \cite{gui14}). The stress tensor $\sigma_{ij}$ is generated by the average auxiliary field  $\sigma_0$ and it is automatically in equilibrium as $\partial_{x_i}P^T_{ij}=0$. $P^T_{ij}$ is the transversal projector, and $\omega_n=2\pi n/\beta$ and $\omega'_n=2\pi (n+\frac{1}{2})/\beta$ are the bosonic and fermionic Matsubara frequencies. The matrix Green function appearing in \eqref{eq1} satisfies 
\begin{eqnarray} 
\int_{x''} \!
\left(\begin{array}{cc}
K_0& -g\\
-g & \hat{V}\\
\end{array}\right)_{xx''}  \!
\left(\begin{array}{cc}
K_0& -g\\
-g & \hat{V}\\
\end{array}\right)^{-1}_{x''x'}\!\!
\equiv
\int_{x''} \!
\left(\begin{array}{cc}
K_0& -g\\
-g & \hat{V}\\
\end{array}\right)_{xx''}\!\!
\mathcal{G}(x'',x')\!
=\!
\left(\begin{array}{cc}
1& 0\\
0 & 1\\
\end{array}\right)\! \delta(x-x'),
\label{eqn2}
\end{eqnarray}
and $\mathcal{G}(x'',x')$ can be computed from \eqref{eqn2} as
\begin{eqnarray}
\mathcal{G}(x,x')=\frac{1}{4\pi^2}\int_{q} e^{iq(x-x')} \frac{1}{\left(K_0 \hat{V}(q)-g^2\right)}
\left(
\begin{array}{cc}
\hat{V}(q) & g \\
g&K_0
\end{array}
\right)
\equiv
\frac{1}{4\pi^2}\int_{q} e^{iq(x-x')}
\hat{\mathcal{G}}(q).
\label{eq8}
\end{eqnarray}
It is clear from \eqref{eq8} that $\mathcal{G}(x,x')=\mathcal{G}(x-x')$. Using this translation invariant Green's function, we can write \eqref{eq1} as
\begin{eqnarray} 
&& \left(\begin{array}{c}
\!\sigma(x)\!\!\\
\!\alpha(x)\!\!\\
\end{array}\right)\!\! = \nonumber \\
&&  \int_y
\left(\begin{array}{cc}
K_0& -g\\
-g & \hat{V}\\
\end{array}\right)_{xy}
\!\!\left(\begin{array}{c}
\!\! P^T_{ij}\partial_{y_i}h(y)\partial_{y_j}h(y)\!-\!\frac{1}{2\beta}\sum_{\omega_n}\!\!P^T_{ij}\partial_{y_i}\partial_{y_j}\! [\rho\omega^2_n\!+\!\kappa\Delta_z^2\!-\!P^T_{lm}\sigma(z)\partial_{z_l}\!\partial_{z_m}]^{-1}_{yy}\!\! \\
\frac{4}{\beta}\sum_{\omega'_n}(i\omega'_n)[(i\omega'_n-\alpha(z)+\mu)^2+v_F^2\nabla_z^2]^{-1}_{yy} \\
\end{array}\right) \label{eq111}
\end{eqnarray}
Henceforth from this equation we suppress the subscript 0 in all average quantities.

Let us now set $\mu=0$ as in \cite{gui14}. It is important to note that the flat membrane $h(x)=0$ is always a solution of (\ref{eq2}). However, the right side of \eqref{eq111} is not zero for $h(x)=0$ and therefore the trivial solution ($h=\sigma=\alpha=0$) is not a valid solution of the SPEs. As we will see in the next section, $h=0$ and non-zero $\sigma$ and $\alpha$ is a valid solution. Nonzero auxiliary fields $\sigma$ and $\alpha$ provide a residual stress that induces plate buckling and rippling below a certain critical temperature and over a critical plate size. In a mathematically related phenomenon, a growing bacterial biofilm can be modeled as a plate with a growth tensor that modifies the elastic part of strain, acts as residual stress and may trigger ripples (called wrinkles in that application) \cite{esp15}. The critical growth term may then be sought by solving an appropriate eigenvalue problem for $h(x)$ coming from the linearized plate equations \cite{der08}.

\section{Flat membrane with constant auxiliary fields}
\label{constant_s_and_a}
We now seek simple solutions of equations \eqref{eq2}-\eqref{eq2c} and \eqref{eq111} with $\mu=0$, $\sigma(x) =\sigma$ (a constant), $\alpha(x)=\alpha$ (a constant) and $h(x)=0$. Using that
\begin{equation}
P^T_{ij}\partial_{x_i}\partial_{x_j}=0, \quad [(i\omega'_n-\alpha(y))^2+v_F^2\nabla_y^2]^{-1}_{xx}=\int_q\frac{1}{(i\omega'_n-\alpha)^2-v_F^2q^2},
\end{equation}
and \eqref{eq8}, Equation (\ref{eq111}) now becomes 
\begin{eqnarray} 
 \left(\begin{array}{c}
\sigma\\
\alpha\\
\end{array}\right) =\left(\begin{array}{cc}
K_0& -g\\
-g & \hat{V}(0)\\
\end{array}\right)\left(\begin{array}{c}
0\\
\frac{4}{\beta}\sum_{\omega'_n}\int_q\frac{i\omega'_n}{(i\omega'_n-\alpha)^2-v_F^2q^2} \\
\end{array}\right)\!, \label{eq13}
\end{eqnarray}
where $\hat{V}(0)\approx V(q\!=\!0)=2\pi e^2R/\epsilon_0$ ($1/R$ is an infrared cutoff resulting from the graphene sheet radius) is its regularized version and the constant appearing in \eqref{eq7} has been neglected. The right side of \eqref{eq13} is computed separately in Appendix \ref{a1} with the  result
\begin{eqnarray}
\frac{4}{\beta}\int_q\sum_{\omega'_n}\frac{i\omega'_n}{(i\omega'_n-\alpha)^2-v_F^2q^2}\sim \frac{v_F\Lambda^2-2\alpha\Lambda}{2\pi v_F}+\frac{\alpha^2}{2\pi v_F^2}-\frac{\pi}{6\beta^2v_F^2}, \label{eq16}
\end{eqnarray}
where $\Lambda=2\pi/a$ is an ultraviolet momentum cutoff ($a$ is the side of a graphene hexagon). The term $\Lambda^2/2\pi$ in \eqref{eq16} is proportional to the area of the Brillouin zone, and it will play a key role in the computation of the critical temperature and radius for the graphene layer. We discuss in Appendix \ref{a2} other possible interpretations of \eqref{eq13} that do not involve setting the chemical potential $\mu=0$ and turn out to be unphysical. From (\ref{eq13}) we obtain
\begin{eqnarray}
\sigma&=& -\frac{\alpha g}{V(0)}, \label{eq17}\\
\frac{\alpha}{V(0)}&\sim & \frac{v_F\Lambda^2-2\alpha\Lambda}{2\pi v_F}+\frac{\alpha^2}{2\pi v_F^2}-\frac{\pi}{6\beta^2v_F^2}. \label{eq18}
\end{eqnarray}
The left side of \eqref{eq18} is much smaller than the right side and it can be ignored, thereby producing the solution  
\begin{eqnarray}
\frac{\alpha}{v_F\Lambda}&\sim & 1\pm \frac{\pi}{\sqrt{3}\beta v_F\Lambda}, \label{eq19}
\end{eqnarray}
which gives $\sigma$ by insertion in (\ref{eq17}).

\section{Linearized equation and eigenvalue problem}
\label{eigenvalue}
We want to ascertain the linear stability of the solution found in the previous section, $h=0$ and constant auxiliary fields. For this we need to know the dynamics and to linearize the corresponding SPEs about this solution. Let us assume that the dynamic SPEs are
\begin{equation}
\mathbf{\mathcal{F}}(\mathbf{u};p)=\mathbf{\mathcal{G}}\!\left(\frac{\partial\mathbf{u}}{\partial t},\frac{\partial^2\mathbf{u}}{\partial t^2}\right)\!,\quad \mathbf{u}=\left(\begin{array}{c}
h\\ \sigma\\ \alpha\\
\end{array}\right)\!,\label{dyn1}
\end{equation}
so that 
\begin{equation}
\mathbf{\mathcal{F}}(\mathbf{u};p)=\mathbf{0},\quad\mathcal{G}(\mathbf{0},\mathbf{0})=\mathbf{0},\label{dyn2}
\end{equation}
are the stationary SPEs (\ref{eq2})-\eqref{eq2c} and (\ref{eq111}). Let $\mathbf{u}_0(p)$ be the flat solution $h=0$ and constant $\sigma(p)$ and $\alpha(p)$ of (\ref{eq17}) and (\ref{eq19}). Here $p$ is a bifurcation parameter, later to be identified as the temperature or the radius of a circular graphene membrane. Linear stability about $\mathbf{u}_0$ follows from substituting $\mathbf{u}(t)=\mathbf{u}_0+e^{\upsilon t}\mathbf{U}$ in (\ref{dyn1}) and keeping terms of order $\mathbf{U}$ in the result. We obtain the eigenvalue problem
\begin{equation}
\frac{\delta\mathbf{\mathcal{F}}}{\delta\mathbf{u}}(\mathbf{u}_0(p);p)\mathbf{U}=\upsilon\!\left[ \frac{\delta\mathbf{\mathcal{G}}}{\delta\partial\mathbf{u}/\partial t}(\mathbf{0},\mathbf{0})+\upsilon \frac{\delta\mathbf{\mathcal{G}}}{\delta\partial^2\mathbf{u}/\partial t^2}(\mathbf{0},\mathbf{0}) \right]\!\mathbf{U},\label{dyn3}
\end{equation}
where $ \frac{\delta\mathbf{\mathcal{F}}}{\delta\mathbf{u}} $, $\frac{\delta\mathbf{\mathcal{G}}}{\delta\partial\mathbf{u}/\partial t}$ and $ \frac{\delta\mathbf{\mathcal{G}}}{\delta\partial^2\mathbf{u}/\partial t^2} $ are functionals acting upon $\mathbf{U}\equiv\mathbf{U}(x)$. For appropriate values of $p$, the real parts of all eigenvalues $\upsilon$ are negative and the flat constant solution $\mathbf{u}_0(p)$ is linearly stable. Then zero eigenvalues or eigenvalues with zero real part give the critical values at which non-flat solutions bifurcate from the flat constant solution. Stationary solutions bifurcate from $\mathbf{u}_0(p)$ at those critical values of $p$ for which $\upsilon=0$. At such $p_c$, (\ref{dyn3}) becomes 
\begin{equation}
\frac{\delta\mathbf{\mathcal{F}}}{\delta\mathbf{u}}(\mathbf{u}_0(p_c);p_c)\mathbf{U}=\mathbf{0}. \label{dyn4}
\end{equation}
While the dynamic SPEs are not known, (\ref{dyn4}) are the {\em known} linearized stationary SPEs about the flat constant solution. We consider (\ref{dyn4}) as an eigenvalue problem for the critical values $p_c$ and proceed to determine ``eigenvalues'' $p_c$. The corresponding eigenvectors $\mathbf{U}$ are buckling modes of the membrane issuing forth from the flat constant solution. When many of these modes become active we may have generated a variety of rippling states of the graphene sheet.

To solve the linearized stationary SPEs, we need appropriate boundary conditions. To solve the linearized plate equation \eqref{eq2}, we need the value of $\sigma_{ij}$ in \eqref{eq2b} for the constant $\sigma$ and $\alpha$ of (\ref{eq17}) and (\ref{eq19}). Now $u=\Delta^{-1}\sigma$ solves the equation $\Delta u= \sigma$ with appropriate boundary conditions. A solution is 
$$u= \frac{\sigma}{4}|\mathbf{x}|^2 + u_0, \quad \Delta u_0=0.$$
 For a circular plate with zero data Dirichlet boundary conditions at $r=R$ the solution is $u=\frac{\sigma}{4}(r^2-R^2)$. Then, by  \eqref{eq2b} and \eqref{eq2c}, $\sigma_{ij}=\sigma\delta_{ij}/2$, and \eqref{eq2} becomes
\begin{eqnarray} 
\Delta^2h=\frac{\sigma}{2\kappa}\Delta h.\label{eq20}
\end{eqnarray}

There are different eigenvalue problems associated to solving Equation \eqref{eq20} with different boundary conditions for a finite graphene membrane. Each of these problems yield its critical temperature and size and allows us to know the shape of different eigenmodes that appear as buckled solutions of the suspended layer of graphene. We consider a finite circular layer of graphene with a free border (natural boundary conditions) or a graphene sheet clamped to a circular hole. Suspended graphene sheets are clamped at their boundaries but the case of a free border is easier to treat mathematically, so we start by analyzing it. 

\subsection{Membrane with free border}
\label{NBC}
The natural boundary conditions to solve \eqref{eq20} for a free circular graphene layer of radius $R$ are
\begin{eqnarray} 
&&\Delta h=0 \quad\mbox{and}\quad \left(\Delta-\frac{\sigma}{2\kappa}\right)\!\partial_r h=0\quad \mbox{at $r=R$.} \label{eq21}
\end{eqnarray}
Let us first solve the eigenvalue problem for $H=\Delta h$: $\Delta H-\frac{\sigma}{2\kappa}H=0$ with $H=0$ at $r=R$. We find the solution
\begin{eqnarray}
H_{n,m}=\frac{1}{a} e^{im\theta}J_m\!\left(\frac{\gamma_{n,m}r}{R}\right)\!,\quad \frac{\sigma}{2\kappa}=-\frac{\gamma_{n,m}^2}{R^2},\label{eq22}
\end{eqnarray} 
where $J_m(\gamma_{n,m})=0$, $n=1,2,\ldots$, $m=0,1,2,\ldots$. The lowest possible value of $\sigma$ in (\ref{eq22}) corresponds to the first zero $\gamma_{1,0}=2.4048$ of the Bessel function $J_0(x)$. Note that $\gamma_{1,0}<\gamma_{1,1}<\gamma_{1,2}<\gamma_{2,0}<\gamma_{1,3}$. Thus there are two eigenvalues corresponding to azimuthal eigenfunctions ($m=1,2$) between the first two eigenvalues corresponding to radially symmetric eigenfunctions with $m=0$.

The solution of $\Delta h=H$ that satisfies the other boundary condition is 
\begin{eqnarray}
h_{n,0}(r)= \frac{R^2}{a\gamma_{n,0}^2}\left[1-J_0\!\left(\frac{r \gamma_{n,0}}{R}\right)\!\right]\!, \label{eq23}
\end{eqnarray} 
for $m=0$ (normalized so that $h_{n,0}(0)=0$), and
\begin{eqnarray}
h_{n,m}(r,\theta)= -\frac{R^2}{a\gamma_{n,m}^2}e^{im\theta} J_m\!\left(\frac{\gamma_{n,m}r}{R}\right)\!, \label{eq24}
\end{eqnarray} 
for $m>0$. 

\begin{figure}
\centering
\subfigure[$h_{1,0}$]{\includegraphics[width=0.3\textwidth]{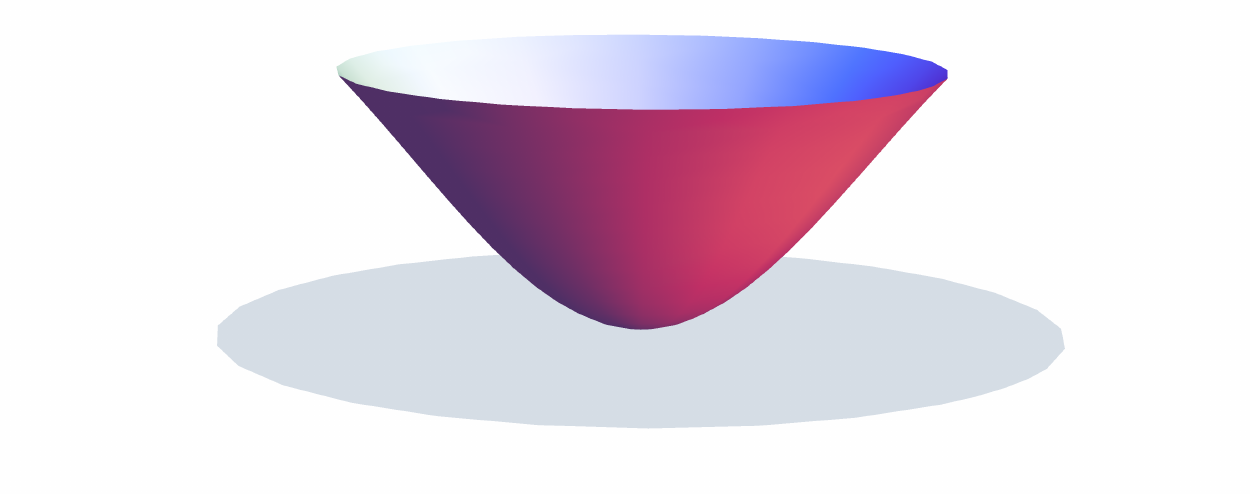}
\label{a}}
\subfigure[$h_{1,1}$]{\includegraphics[width=0.3\textwidth]{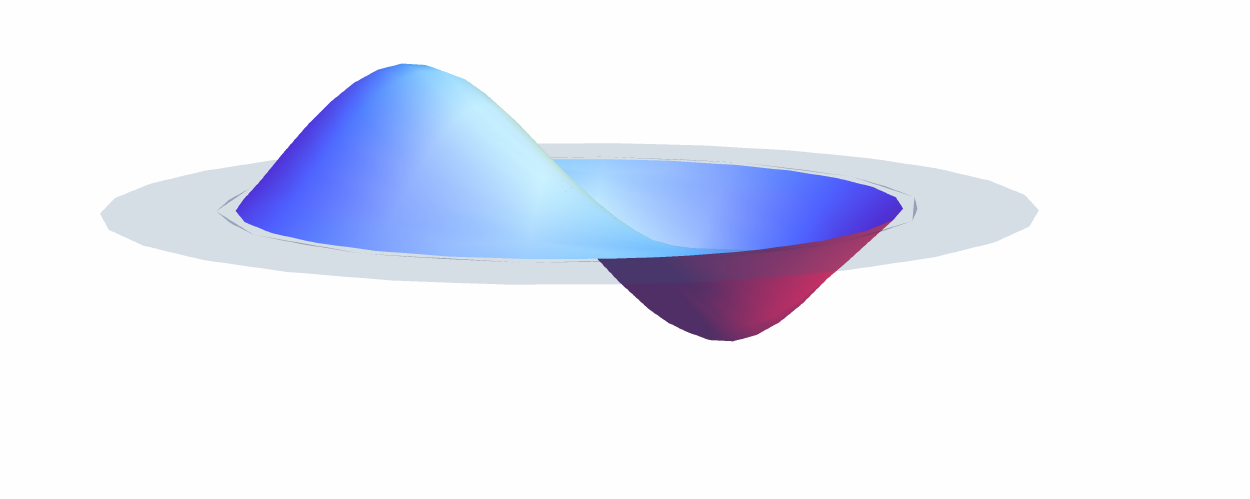}
\label{c}}
\subfigure[$h_{1,2}$]{\includegraphics[width=0.3\textwidth]{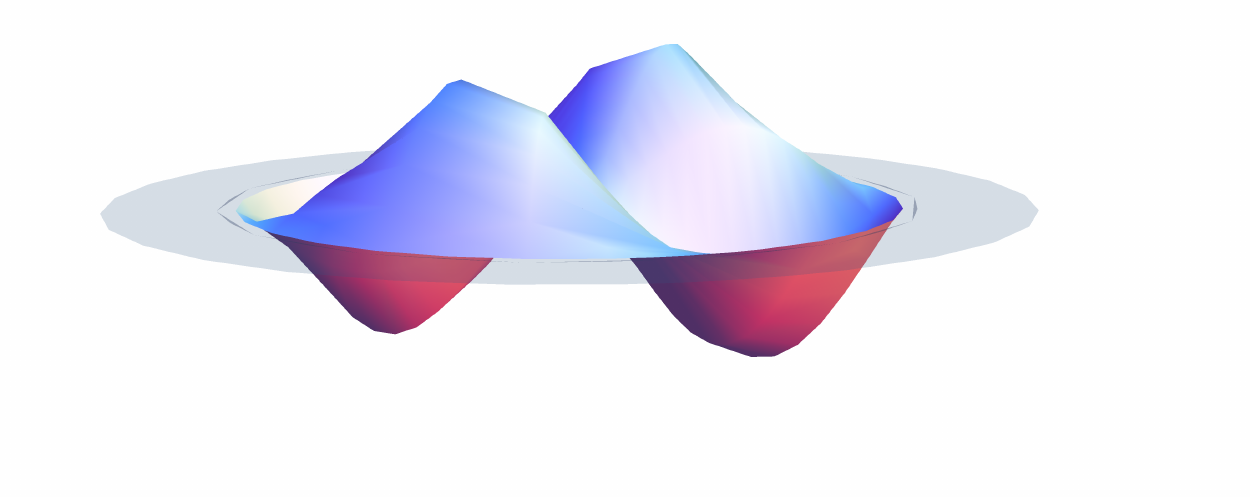}
\label{d}}
\subfigure[$h_{2,0}$]{\includegraphics[width=0.3\textwidth]{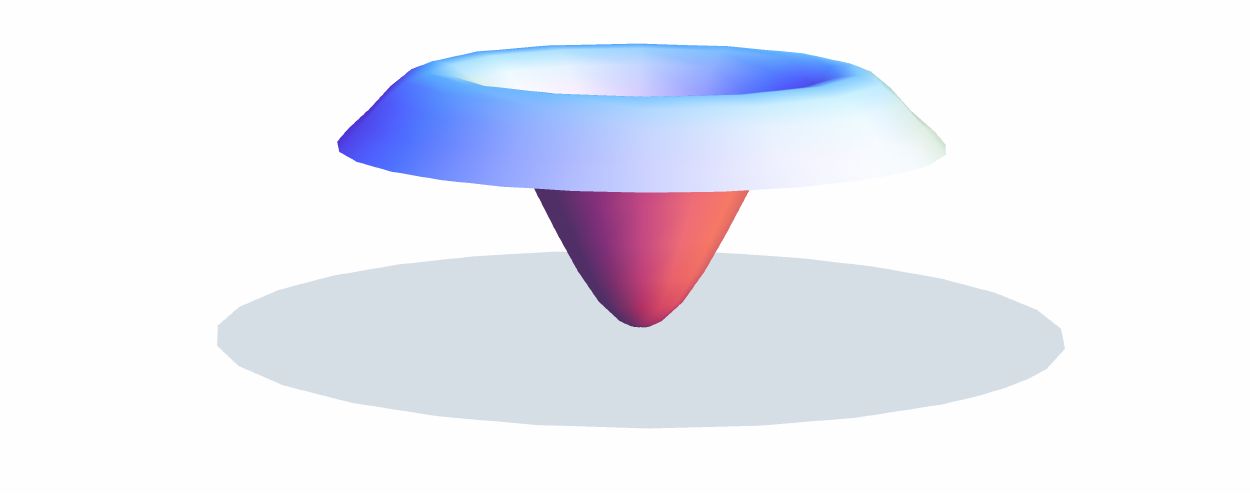}
\label{b}}
\caption{(Color online) Modes appearing over their respective critical radii, see table \ref{table}. For a suspended layer of graphene of radius $R$ and free boundaries, all modes with $R_c<R$ can combine to produce ripples or buckling. The grey disk is the  plane $h=0$.
}
\label{fig1}
\end{figure}

 We are interested in computing the critical temperature and radii. To this end, we use (\ref{eq22}) with equations (\ref{eq17}) and (\ref{eq19}), and find that ripples appear below the critical temperature
\begin{eqnarray}
T_c= \frac{2\sqrt{3} v_F}{a}\left|1- \frac{e^2}{\epsilon_0 v_F}\,\frac{2a\kappa}{Rg}\gamma_{1,0}^2\right|\!. \label{eq25}
\end{eqnarray}
Note that the critical temperature decreases as the size $R$ decreases. As $R/a\to\infty$, the critical temperature tends to $\frac{2\sqrt{3} v_F}{a}\approx 16.05$ eV (about 186,000 K). For finite membrane size, (\ref{eq25}) with $\gamma_{n,m}$ instead of $\gamma_{1,0}$ produces the following numerical estimate,
\begin{equation}
1 - 5.4\times 10^{-6}\, T_{n,m} = \frac{1.6\,\gamma_{n,m}^2}{R}. \label{eqn3}
\end{equation}
Here $T_{n,m}$ is the temperature (measured in Kelvin) at which the mode $h_{n,m}$ appears and $R$ is measured in nanometers. The values considered for the parameters are $g=3.9$ eV (the lowest value considered in \cite{gui14}) and $\kappa=1$ eV. At room temperature, equation \eqref{eqn3} gives us different critical radii above which the different modes $h_{n,m}$ (characterised by $\gamma_{n,m}$) appear, see table \ref{table} and figure \ref{fig1}.
\begin{table}[]
\centering
\begin{tabular}{ccccc}\hline
 $h$: & $h_{1,0}$& $h_{1,1}$& $h_{1,2}$&  $h_{2,0}$ \\ \\
 $R_c$ (nm): &$9$ &$24$ & $43$ &$50$  \\ \hline
\end{tabular}
\caption{Critical radii given by Equation \eqref{eqn3} for a circular layer graphene at room temperature with natural boundary conditions. Mode $h_{n,m}$ appears from the flat solution for $R>R_c$. }
\label{table}
\end{table}

\subsection{Clamped Membrane}
\label{CM}
In this more realistic case, the boundary conditions are $\tilde{h}=0$ and $\partial_r\tilde{h}=0$ at $r=R$ instead of (\ref{eq21}). We use $\tilde{h}$ instead of $h$ to distinguish the out-of-plane displacement in the clamped case. Following the same procedure as for equations \eqref{eq22} and \eqref{eq23}, we get 
\begin{eqnarray}
H_{n,m}=\frac{1}{a} e^{im\theta}J_m(kr),\quad k^2=-\frac{\sigma}{2\kappa},\label{eq26}
\end{eqnarray} 
instead of \eqref{eq22}, and in which $k$ is not yet determined. Now the radial part $\tilde{h}_m(r)$ satisfies 
\begin{equation}
\left(\partial_r^2+\frac{1}{r}\partial_r-\frac{m^2}{r^2}\right)\!\tilde{h}_m=\frac{1}{a}J_m(kr). \label{eq221}
\end{equation}
For $m=0$, this equation is 
$$\partial_r(r\partial_r\tilde{h})=\frac{r}{a}J_0(kr)\Longrightarrow r\partial_r\tilde{h}=\frac{1}{ka}[r J_1(kr) -RJ_1(kR)],$$
in which we have used the boundary condition $\tilde{h}'(R)=0$. Integrating once and using $\tilde{h}(R)=0$, we obtain
\begin{equation}
\tilde{h}(r)=\frac{1}{k^2a}[J_0(kR)-J_0(kr)]-\frac{R}{ka}\ln\!\left(\frac{r}{R}\right) J_1(kR).\nonumber
\end{equation}
The vertical displacement is unbounded at $r=0$ unless 
\begin{equation}
J_1(kR)=0 \Longrightarrow \frac{\sigma}{2\kappa}=-\frac{\gamma_{n,1}^2}{R^2}. \label{eq224}
\end{equation}
Then
\begin{equation}
\tilde{h}_{n,0}(r)=\frac{R^2}{\gamma_{n,1}^2a}\!\left[J_0(\gamma_{n,1})-J_0\!\left(\frac{\gamma_{n,1}r}{R}\right)\!\right]\!. \label{eq222}
\end{equation}
Similarly, for $m>0$ the solution of $\Delta \tilde{h}=H_{n,m}$ is 
\begin{eqnarray}
\tilde{h}_{n,m}(r,\theta)=\left(\frac{c_1}{r^m}+c_2r^m\right)\!e^{i m\theta}+\frac{r^m}{a} e^{i m\theta}\int_0^r \frac{dr}{r^{2m+1}}\int_0^r ds\, s^{m+1}J_m(ks). \label{eq27}
\end{eqnarray} 
Clamped boundary conditions yield
\begin{eqnarray}
&&c_2=-\frac{1}{a} \int_0^R \frac{dr}{r^{2m+1}}\int_0^r ds\, s^{m+1}J_m(ks)-\frac{c_1}{R^{2m}}, \label{eq28}\\
&&c_1=\frac{1}{2ma}\int_0^Rds\, s^{m+1}J_m(ks)=\frac{R^{m+2}}{2ma}\int_0^1ds\, s^{m+1}J_m(kRs)=\frac{R^{m+1}}{2mka}J_{m+1}(kR).\label{eq29}
\end{eqnarray} 
The condition that $\tilde{h}_{n,m}$ be bounded at $r=0$ produces $c_1=0$. Thus $J_{m+1}(kR)=0$, i.e., $k=\gamma_{n,m+1}/R$ and therefore
\begin{eqnarray}
\frac{\sigma}{2\kappa}=-\frac{\gamma_{n,m+1}^2}{R^2}.\label{eq30}
\end{eqnarray} 
Equation \eqref{eq27} becomes
\begin{eqnarray}
\tilde{h}_{n,m}(r,\theta)&=&\frac{r^m}{a} e^{i m\theta}\int_R^r \frac{dr}{r^{2m+1}}\int_0^r ds\, s^{m+1}J_m\!\left(\frac{\gamma_{n,m+1}s}{R}\right)\! = -\frac{R^2e^{i m\theta}}{\gamma_{n,m+1}a}\!\left(\frac{r}{R}\right)^{m}\!\int_{r/R}^1\frac{ds}{s^{m}} J_{m+1}(\gamma_{n,m+1}s)\nonumber\\
&=& \frac{R^2 e^{i m\theta}}{a\gamma_{n,m+1}^2}\!\left\{\left(\frac{r}{R}\right)^m J_m(\gamma_{n,m+1})-J_m\!\left(\gamma_{n,m+1}\frac{r}{R}\right)\!\right\}\!, \label{eq271}
\end{eqnarray} 
which agrees with \eqref{eq222} for $m=0$.

\begin{figure}
\centering
\subfigure[ $\tilde{h}_{1,0}$]{\includegraphics[width=0.3\textwidth]{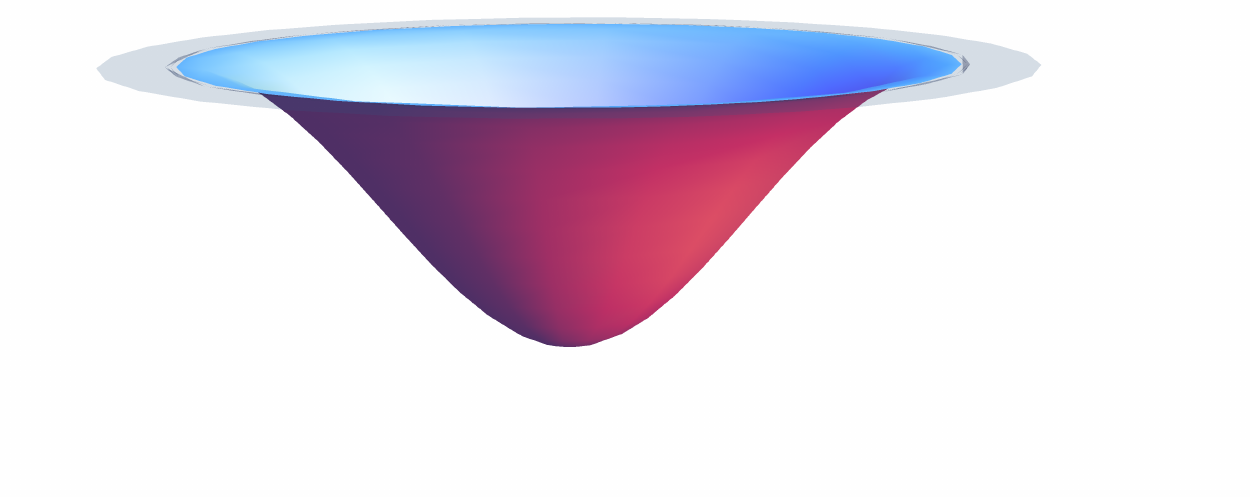}
\label{a22}}
\subfigure[ $\tilde{h}_{1,1}$]{\includegraphics[width=0.3\textwidth]{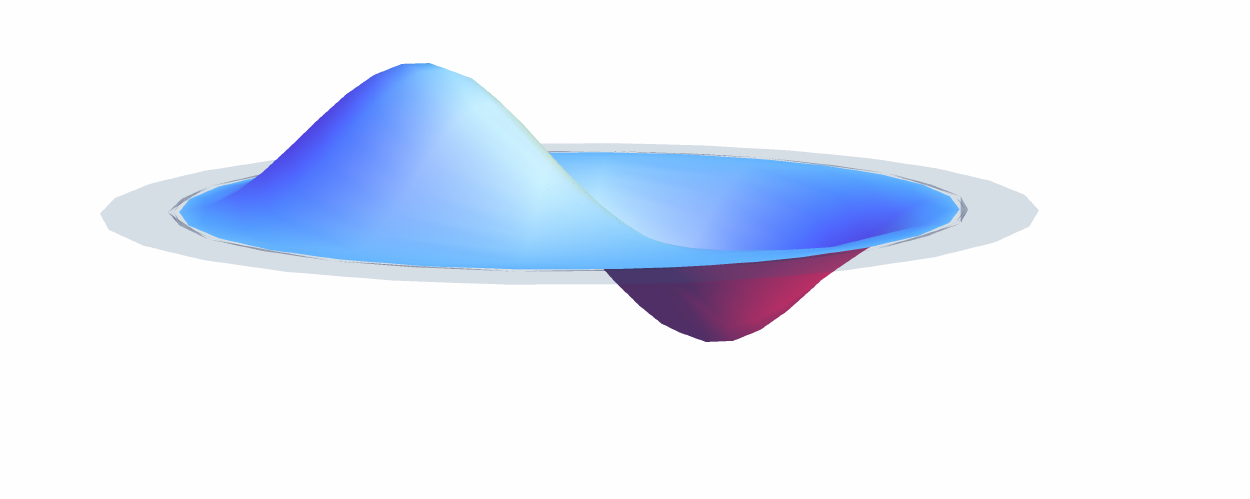}
\label{d2}}
\subfigure[ $\tilde{h}_{1,1}$]{\includegraphics[width=0.3\textwidth]{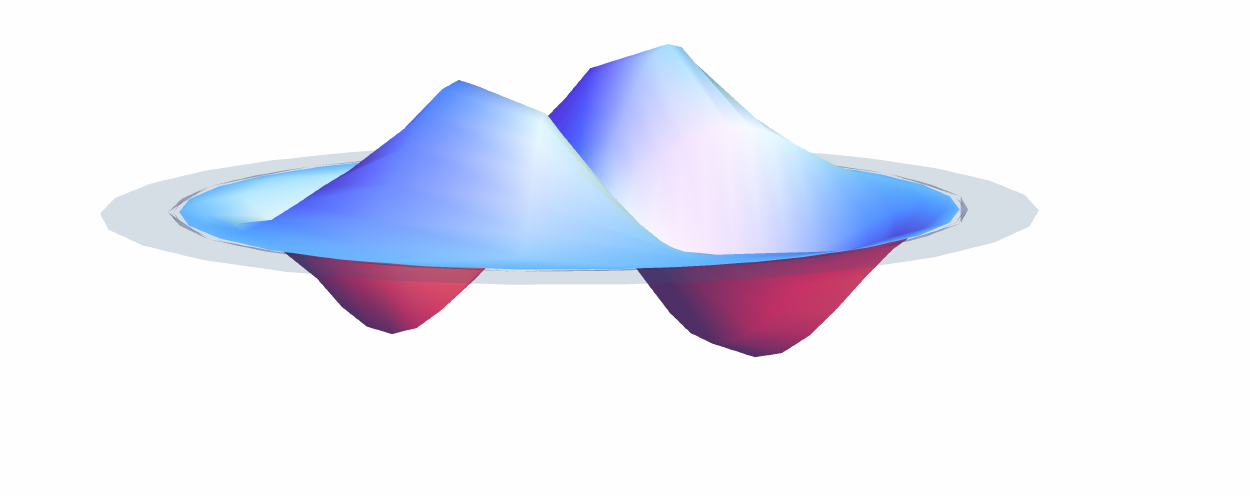}
\label{b2}}
\subfigure[ $\tilde{h}_{2,0}$]{\includegraphics[width=0.3\textwidth]{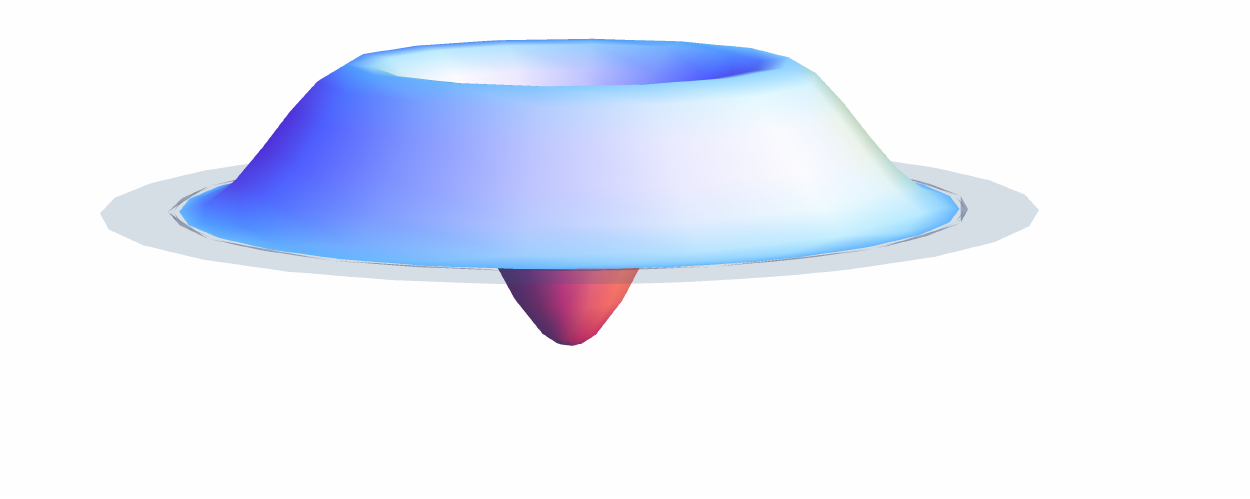}
\label{c2}}
\caption{(Color online) Modes appearing over their respective critical radii, see table \ref{table}. In a suspended layer of graphene clamped on a circular hole of radius $R$ all modes with $R_c<R$ can appear and combine to form ripples and buckling. The grey disk  is the substrate plane $h=0$.
}
\label{fig2}
\end{figure}

According to \eqref{eq30}, the critical radii are given by \eqref{eqn3} with $\gamma_{n,m+1}$ instead of $\gamma_{n,m}$. Then the critical radii of table \ref{table2} are greater than those in table \ref{table}, and buckling of a clamped graphene membrane should be observable only for radii over $24$ nm. Thus a clamped membrane remains flat for larger radii than in the case of natural free boundary conditions. This is according to our intuition that it is harder to buckle a clamped membrane than a membrane with a free border. The lowest possible value of $\sigma$ in (\ref{eq22}) corresponds to $\gamma_{1,1}=3.8317$. Note that $\gamma_{1,1}<\gamma_{1,2}<\gamma_{1,3}<\gamma_{2,1}$. Again there are two eigenvalues corresponding to azimuthal eigenfunctions ($m=1,2$) between the first two eigenvalues corresponding to radially symmetric eigenfunctions with $m=0$. These first four modes given by \eqref{eq271} are depicted in figure \ref{fig2}.
\begin{table}[]
\centering
\begin{tabular}{ccccc}\hline\\
 $\tilde{h}$: & $\tilde{h}_{1,0}$&$\tilde{h}_{1,1}$&$\tilde{h}_{1,2}$& $\tilde{h}_{2,0}$\\ \\
 $R_c$ (nm): &$24$ &43 &66 &80  \\
 \hline
 \end{tabular}
\caption{Critical radii given by Equation \eqref{eqn3} for a circular layer graphene at room temperature with clamped boundary conditions. Mode $\tilde{h}_{n,m}$ appears from the flat solution for $R>R_c$.}
\label{table2}
\end{table}

\section{Conclusions}
\label{conc}
In conclusion, the stationary saddle-point equations give the vertical displacement of a graphene membrane and auxiliary fields associated to curvature and to charge fluctuations, provided elasticity and phonon-electron interaction are considered \cite{gui14}. We have solved these equations for a flat circular graphene sheet under constant auxiliary fields. We have then solved the SPEs linearized about the flat membrane solution and subject to either free or clamped boundary conditions. The latter  problem has nonzero solutions only for discrete values of the auxiliary fields. These discrete values correspond to critical values of temperature and membrane radius at which buckling membrane solutions issue forth from the flat membrane. The flat membrane is stable only for temperatures larger than critical and radii smaller than critical. For an infinite membrane, the predicted critical temperature is extremely large and unphysical  (186,000 K). However, critical radii at room temperature are in the nanometer range. Different modes of membrane buckling appear below specific critical temperatures and over specific critical radii, see tables \ref{table} and \ref{table2}. The stability of these modes is to be analytically determined, but experiments describing rippling \cite{mey07} and buckling \cite{schoelz15} of graphene show that the flat solution is not stable in most cases: Flat graphene has only been observed when it stands over a substrate \cite{lui09}. 

The calculated critical radii (tables \ref{table} and \ref{table2}) suggest that the different modes appearing in figures \ref{fig1} and \ref{fig2} could be observed for a specific temperature and membrane size. This brings the opportunity to develop new experiments, as no buckling should appear below the minimum critical radius $R_c$. Similar experimental procedures to the ones in \cite{lin12,svensson11,xu14}, where suspended graphene layers were forced to buckle using an external electrical potential or a STM, could be used to determine the value of $R_c$ and the validity of our results. For this purpose, suspended graphene layers may be clamped to holes of different dimensions, under different temperatures. Checking when the layer may or not buckle, it should be  possible to get a better estimation of $g$ and $\kappa$ by fitting the results of experiments with equation \eqref{eq25}.

A different approach to study the effects due to the existence of non-flat modes would be experiments with graphene resonators \cite{bunch07,eichler11}. In this setting, when the external driving force is sufficiently strong to deform the graphene layer close to one of the stable buckling  modes (away from the flat configuration), the response of the resonator should be extremely nonlinear. Instead of oscillating around a single potential well (more or less parabolic) corresponding to the rippled configuration, the system may oscillate under influence of two different potential wells (the buckling modes). This effect could be measured by observing oscillation damping once the driving force is turned off.

Characterization of ripples in suspended graphene is not yet accurate enough to compare with our results. Once the experimental techniques have been improved, experiments with suspended graphene layers clamped to holes of different radii should show ripples that are combination of the possible bifurcating modes for these membranes (table \ref{table2}).

The study of different solutions of the SPEs with position-dependent auxiliary fields $\sigma$ and $\alpha$ is left as future work. From a theoretical point of view, it is also interesting to take into account the influence of external forces, such as an electrostatic force, and to study the resulting dynamical effects. Then we may be able to reproduce the transition from the flat membrane with superimposed ripples to the buckled membrane and compare with existing STM experimental results \cite{schoelz15}.

\acknowledgments
This work has been supported by the Ministerio de Econom\'\i a y Competitividad grants
FIS2011-28838-C02-01 and MTM2014-56948-C02-02. MRG also acknowledges support from Ministerio de Educaci\'on, Cultura y Deporte through the FPU program grant FPU13/02971. This work was partly carried out during a stay of both authors at University of Iceland. We thank the University of Iceland for hospitality. LLB acknowledges support by a grant from Iceland, Liechtenstein and Norway through the EEA Financial Mechanism operated by Universidad Complutense de Madrid. 

\appendix
 \section{Matsubara frequency sum}\label{a1}
The sum over Matsubara frequencies in the expression on the right side of equation \eqref{eq13} gives \cite{nie95}
\begin{eqnarray}\nonumber
&&\frac{1}{\beta}\sum_{\omega'_n}\frac{i\omega'_n}{(i\omega'_n-\alpha)^2-v_F^2q^2}=\frac{1}{2\beta}\sum_{\omega'_n}\left(\frac{1}{i\omega'_n-\alpha-v_Fq}+\frac{1}{i\omega'_n-\alpha+v_Fq}+\frac{2\alpha}{(i\omega'_n-\alpha)^2-v_F^2q^2}\right)\\
&&=\frac{1}{2}\left(\frac{1}{1+ e^{\beta(\alpha+v_Fq)}}+\frac{1}{1+e^{\beta(\alpha-v_Fq)}}\right)\!+\frac{\alpha}{2qv_F}\!\left(\frac{1}{1+ e^{\beta(\alpha+v_Fq)}}-\frac{1}{1+ e^{\beta(\alpha-v_Fq)}}\right)\!=\frac{1}{2qv_F}\nonumber\\
&&\times\!\left(\frac{\alpha+v_Fq}{1+ e^{\beta(\alpha+v_Fq)}}- \frac{\alpha-v_Fq}{1+ e^{\beta(\alpha-v_Fq)}}\right)\!=\frac{1}{2qv_F}\!\left(\frac{\alpha+v_Fq}{1+ e^{\beta(\alpha+v_Fq)}}- \frac{v_Fq-\alpha}{1+ e^{\beta(v_Fq-\alpha)}}\right)\!+\frac{v_Fq-\alpha}{2qv_F}. \label{eq14}
\end{eqnarray}
Then 
\begin{eqnarray}\nonumber
&&\frac{4}{\beta}\int_q\sum_{\omega'_n}\frac{i\omega'_n}{(i\omega'_n-\alpha)^2-v_F^2q^2}=\frac{1}{\pi v_F}\int_0^\Lambda \!\left(\frac{\alpha+v_Fq}{1+ e^{\beta(\alpha+v_Fq)}}- \frac{v_Fq-\alpha}{1+ e^{\beta(v_Fq-\alpha)}}\right)\! dq\\
&&+\!\frac{1}{\pi}\!\int_0^\Lambda\!\!\left(q-\frac{\alpha}{v_F}\right)\!dq\!=\! \frac{v_F\Lambda^2\!-2\alpha\Lambda}{2\pi v_F}\!-\!\frac{1}{\pi\beta^2v_F^2}\!\left(\int_{e^{-\beta(v_F\Lambda+\alpha)}}^{e^{-\beta\alpha}}\!\frac{\ln t\, dt}{1+t}-\!\int_{e^{-\beta(v_F\Lambda-\alpha)}}^{e^{\beta\alpha}}\!\!\frac{\ln t\, dt}{1+t}\!\right)\! \label{eq15}
\end{eqnarray}
We now split the last integral on the right hand side of (\ref{eq15}) as
\begin{eqnarray}\nonumber
\int_{e^{-\beta(v_F\Lambda-\alpha)}}^{e^{\beta\alpha}}\frac{\ln t\, dt}{1+t}\!&=&\!\left(\int_{e^{-\beta(v_F\Lambda-\alpha)}}^1+\int_1^{e^{\beta\alpha}}\right)\frac{\ln t\, dt}{1+t}=\int_{e^{-\beta(v_F\Lambda-\alpha)}}^1\frac{\ln t\, dt}{1+t}-\int_1^{e^{\beta\alpha}}\frac{\ln t\, dt}{(1+t)t}+\left.\frac{1}{2}(\ln t)^2\right|_1^{e^{\beta\alpha}}\\
& \sim& \frac{\alpha^2\beta^2}{2}+\int_{0}^1\frac{\ln t\, dt}{1+t}- \int_1^{\infty}\frac{\ln t\, dt}{(1+t)t}=\frac{\alpha^2\beta^2}{2}-\frac{\pi^2}{6}.\nonumber
\end{eqnarray}
The result is 
\begin{eqnarray}
&&\frac{4}{\beta}\int_q\sum_{\omega'_n}\frac{i\omega'_n}{(i\omega'_n-\alpha)^2-v_F^2q^2}\sim \frac{v_F\Lambda^2-2\alpha\Lambda}{2\pi v_F}+\frac{\alpha^2}{2\pi v_F^2}-\frac{\pi}{6\beta^2v_F^2},
\end{eqnarray}
because the other integrals are of order $O(e^{-\beta\alpha})\ll 1$. In these computations $q$ is upper bounded by $\Lambda=2\pi/a$.

\section{Selecting the chemical potential} \label{a2}
Keeping a nonzero chemical potential, \eqref{eq13} and \eqref{eq16} yield 
\begin{eqnarray}
\sigma=-\frac{g\alpha}{V(0)},\,\,\frac{\alpha}{V(0)}= \frac{4}{\beta}\int_q\sum_{\omega'_n}\frac{i\omega'_n}{(i\omega'_n-\alpha+\mu)^2-v_F^2q^2}\sim   \frac{(\alpha-\mu-v_F\Lambda)^2-\frac{\pi^2}{3\beta^2}}{2\pi v_F}.
\label{b1}
\end{eqnarray}
These are two equations for three unknowns, $\sigma$, $\alpha$ and $\mu$, so that we can give a value to $\mu$ and solve for $\sigma$ and $\alpha$. We have already seen the consequences of setting $\mu=0$. What do we get from other choices?

1. Assume that $\mu$ is selected so that $\mu+v_F\Lambda=0$ and therefore the right hand side of \eqref{b1} (which represents the Fermionic propagator in real space for $x=0$) becomes finite as $\Lambda\to\infty$. Following the same steps as in previous sections, e.g. using equations \eqref{eq18} and \eqref{eq22}, we get (the minus sign corresponds to $g<0$) 
\begin{equation}
\frac{8\kappa\pi e^4}{gv_F^2\epsilon_0^2}\gamma_{nm}=1\pm\sqrt{1+\frac{2\pi^2e^4}{9\epsilon_0^2v_F^4\beta^2}\frac{R^2}{\beta^2}}.
\end{equation}
The right hand side of this equation contains the product $R/\beta=RT$. This produces a critical radius (at which the graphene sheet starts buckling) that decreases when the temperature increases. This is unphysical because increasing temperature would favor rippling, contrary to experimental evidence \cite{bao09}.

2. Let us set $R=\infty$ and $\alpha=0$ in \eqref{b1}, thereby obtaining
\begin{equation}
\mu+v_F\Lambda=\pm \frac{\pi}{\sqrt{3}\beta}.
\end{equation}
If $\alpha$ is no longer zero for finite radius, \eqref{eq18} gives
\begin{equation}
\alpha\left(\alpha\mp \frac{2\pi}{\sqrt{3}\beta}-\frac{2\pi v_F^2}{V(0)}\right)=0.
\end{equation}
Then either $\alpha=0$ and $\sigma=0$, which contradicts \eqref{eq22}, or 
\begin{equation}
\alpha=2\pi \left(   \pm \frac{1}{\sqrt{3}\beta} +\frac{v_F^2}{V(0)}  \right)\!.
\end{equation}
In this case, the same arguments based on Equations \eqref{eq18} and \eqref{eq22} provide
\begin{equation}
\gamma^2_{nm}=\frac{g\pi v_F^2\epsilon_0^2}{\kappa e^4}\pm \frac{g\pi\epsilon_0}{\sqrt{3}\kappa e^2} \frac{R}{\beta}.
\label{b5}
\end{equation}
Once the numerical values of the constants are taken into account, we notice that the left hand side of \eqref{b5} is positive, whereas the second term in the right hand side of this equation (the minus sign corresponds to $g<0$) is proportional to $T R$, and this again produces the same unphysical effect as before.

Thus we conclude that the physically meaningful choice of chemical potential is $\mu=0$ as in \cite{gui14}. We do not renormalize the parameters of the model and leave the natural ultraviolet and infrared cutoffs $\Lambda=2\pi/a$ and $q_0=2\pi/R$, respectively, in terms of the lattice constant $a$ and the radius $R$. 

\end{document}